\documentclass[10pt,aps,prl,twocolumn,groupedaddress,showkeys]{revtex4-2}
\usepackage{amsmath}
\usepackage{graphicx}
\usepackage[colorlinks=true,linkcolor=blue,urlcolor=blue,citecolor=blue,anchorcolor=blue]{hyperref}
\DeclareGraphicsExtensions{.pdf,.eps,.png,.jpg,.mps}
\usepackage{epstopdf}
\usepackage{threeparttable}
\usepackage{color}

\begin{document}

\title{Hysteresis-Free High Mobility Graphene Encapsulated in Tungsten Disulfide}
\author{Karuppasamy Pandian Soundarapandian$^1$}
\author{Domenico De Fazio$^{1,2}$}
\email[]{domenico.defazio@unive.it}
\author{Sefaattin Tongay$^3$}
\author{Frank H. L. Koppens$^1$}
\email[]{frank.koppens@icfo.eu}
\affiliation{$^1$ICFO-Institut de Ciencies Fotoniques, The Barcelona Institute of Science and Technology, 08860 Castelldefels (Barcelona), Spain}
\affiliation{$^2$Department of Molecular Sciences and Nanosystems, Ca’ Foscari University of Venice, 30172 Venezia (Veneto), Italy}
\affiliation{$^3$School for Engineering of Matter, Transport and Energy, Arizona State University, 85287 Tempe (Arizona), USA}

\keywords{Graphene, transition metal dichalcogenides, encapsulation, hysteresis, mobility}

\begin{abstract}

High mobility is a crucial requirement for a large variety of electronic device applications. The state-of-the-art for high quality graphene devices is based on heterostructures made with graphene encapsulated in $>80\,$nm-thick flakes of hexagonal boron nitride (hBN). Unfortunately, scaling up multilayer hBN while precisely controlling the number of layers remains an elusive challenge, resulting in a rough material unable to enhance the mobility of graphene. This leads to the pursuit of alternative, scalable materials, which can be simultaneously used as substrate and encapsulant for graphene. Tungsten disulfide (WS$_2$) is a transition metal dichalcogenide, which was successfully grown in large ($\sim$mm-size) multi-layers by chemical vapour deposition. However, the resistance \textit{vs} gate voltage characteristics when gating graphene through WS$_2$ exhibit largely hysteretic shifts of the charge neutrality point (CNP) in the order of $\Delta n\sim$2.6$\cdot$10$^{11}$\,cm$^{-2}$, hindering the use of WS$_2$ as a reliable encapsulant. The hysteresis originates due to the charge traps from sulfur vacancies present in WS$_2$. In this work, we report for the first time the use of WS$_2$ as a substrate and the overcoming of hysteresis issues by chemically treating WS$_2$ with a super-acid, which passivates these vacancies and strips the surface from contaminants. The hysteresis is significantly reduced below the noise level by at least a factor five (to $\Delta n<$5$\cdot$10$^{10}$\,cm$^{-2}$) and, simultaneously, the room-temperature mobility of WS$_2$-encapsulated graphene is as high as $\sim$6.2$\cdot$10$^{4}$\,cm$^{-2}$V$^{-1}$s$^{-1}$ at a carrier density $n$ $\sim$1$\cdot 10^{12} $\,cm$^{-2}$. Our results promote WS$_2$ to a valid alternative to hBN as encapsulant for high-performance graphene devices. 
\end{abstract}

\maketitle

High mobility is a crucial requirement for multiple electronic and optoelectronic applications such as field effect transistors\cite{Sze2006a}, modulators\cite{Romagnoli2018a}, photodetectors\cite{Kang2008,Muench2019a} and sensors\cite{Wang2016a}. Owing to its ultra-high ($>$10$^{5}$\,cm$^{-2}$V$^{-1}$s$^{-1}$) room-temperature mobility\cite{Wang2013g}, complemented by broadband absorption\cite{Nair2008b}, scalability\cite{Bae2010a,Polsen2015} and compatibility to the complementary metal-oxide-semiconductor (CMOS) platform\cite{Goossens2017b}, graphene is quickly rising for consideration in advanced multi-purpose technologies\cite{Akinwande2019,Romagnoli2018a}. The room-temperature mobility in graphene is limited by the scattering of carriers with acoustic phonons\cite{Hwang2008b} and it is typically inversely proportional to the residual charge carrier density ($n$*)\cite{Couto2014}, which arises from local strain fluctuations in graphene\cite{Couto2014,Neumann2015}. In this context, the choice of the perfect substrate, dictated by the need of atomic flatness and absence of charge traps, plays a vital role for preserving the extra-ordinary properties of graphene\cite{Dean2010b}.

Suspended graphene exhibits extremely high mobility ($\sim$2$\cdot$10$^{5}$\,cm$^{-2}$V$^{-1}$s$^{-1}$)\cite{BolotinSSC2008} but the integration of such devices is impractical\cite{Giambra2021}. Several substrates such as aluminium dioxide (Al$_2$O$_3$)\cite{KimAPL2009}, aluminium nitride (AlN)\cite{OhAPL2014}, fused silica\cite{RamonIEEE2012}, strontium titanate (SrTiO$_3$)\cite{CoutoPRL2011} and calcium fluoride (CaF$_2$)\cite{KlarPRB2013} were tested to preserve the properties of suspended graphene. Nevertheless, hexagonal boron nitride (hBN), an atomically flat, layered dielectric material is so far the unrivalled choice for encapsulating graphene\cite{PurdieNC2018}. State-of-the-art (SOTA) high-quality graphene (Gr) devices are encapsulated with hBN flakes and assembled through a novel stamping and cleaning technique\cite{HuangNC2020}. The highest values of mobility measured in such hBN/Gr/hBN heterostructure at room temperature exceeds $\sim$10$^5$\,cm$^2$V$^{-1}$s$^{-1}$ at a carrier density of $\sim$10$^{12}$\,cm$^{-2}$ with a $n$* of  $2\cdot$10$^{9}$\,cm$^{-2}$ \cite{HuangNC2020, Pezzini2020}. In such experiments atomically-flat multilayer flakes of hBN $>80$\,nm-thick were used both as a substrate and as a capping layer, in order to completely screen charge traps and roughness from the substrate underneath, but also to protect graphene from exposure to air-contaminants\cite{HuangNC2020}. However, growing multilayer hBN films with equal flatness to the exfoliated counterpart, and being able to precisely control the number of layers, remains an open challenge\cite{SMKimNC2015,ShenAM2021}. Additionally, hBN grown by chemical vapour deposition (CVD) tends to possess a high concentration of point defects, wrinkles and grain boundaries, resulting in a significant increase of the overall roughness\cite{ShenAM2021}. Consequently, multilayer CVD-grown hBN does not represent at the moment an ideal candidate for scalable encapsulation of high quality graphene based devices\cite{Shautsova2016,Giambra2021}

Ultra-high room temperature mobility ($>$2.5$\cdot$10$^5$\,cm$^2$V$^{-1}$s$^{-1}$) at a carrier density of ($\sim$10$^{12}$\,cm$^{-2}$) was reported when using TMD flakes as capping layers while still using hBN as a substrate\cite{LucasARXV2019}. Ref.\citenum{LucasARXV2019} claimed that this increase in mobility with respect to fully-hBN-encapsulated Gr might be due to a modification of the acoustic phonon bands in Gr, although the origin of this mechanism has yet to be understood. Ref.\citenum{Lucas2DM2016} performed Raman characterization of hBN/Gr heterostructures placed on a variety of substrates. The 2D peak in the Raman spectrum of graphene originates from a double-resonant process and it is the most intense measurable feature\cite{FerrPRL2006,FerrNN2013}. The full-width-at-half-maximum (FWHM) of the 2D peak (FWHM(2D)) has been shown to be related to the amount of nanometre-scale strain variations in the sample\cite{Neumann2015}. hBN/Gr heterostructures with TMDs as a substrate measured in Ref.\citenum{Lucas2DM2016} exhibited the smallest values of FWHM(2D), with values similar to the ones obtained with full hBN encapsulation\cite{Lucas2DM2016}.  
Refs.\citenum{JungSR2021,CunninghamJPC2016,SchramESSDERC2017} reported growing uniform multilayer transition metal dichalcogenides (TMDs) by CVD, as well as demonstrating the possibility to integrate these materials in the CMOS back-end-of-line (BEOL)\cite{SchramESSDERC2017}. Thus, TMDs are potentially promising substrates for Gr. 

One potential disadvantage of using TMDs as a substrate, and possibly as a gate dielectric in Gr-based devices, is their proneness to hysteretic behavior\cite{HWang2018,HAn2020}. Hysteresis can be calculated by extracting the difference in charge carrier density ($\Delta n$) at the charge neutrality point (CNP) between the forward and reverse sweeps\cite{WangACN2010}, with the CNP being the transition between electron and hole doping in the graphene channel\cite{NovoS2004}. The origin of the hysteresis in TMDs can be explained by the presence of trap states in the TMD layer\cite{RohJID2016}. Residues from the fabrication, active absorption of molecules at atomic vacancies and other defect sites are all possible forms of trap states\cite{LanNR2020}. A high density of defect sites $\sim~10^{13}$\,cm$^{-2}$\cite{QiuNC2013} not only degrades the electrical performance of the material but also represents a detrimental factor in the photoluminescence brightness of semiconducting TMDs by creating channels for defect-mediated non-radiative recombination processes\cite{AmaniS2015}. Ref.\citenum{AmaniS2015} showed that a chemical passivation of the defects in sulfur-based monolayer TMDs, obtained by using the super-acid bis(trifluoromethane) sulfonimide (TFSI), leads to a factor of 190-fold improvement in the material photoluminescence intensity. Although the exact passivating mechanism is not fully understood, the authors hypothesize a combination of two processes: on the one hand the removal of contaminants, driven by the protonating nature of the acid and, on the other hand, an energy-favourable reconstruction of the sulfur vacancies promoted by hydrogenation and the rich presence of sulfur atoms in the super-acid\cite{AmaniS2015}. Many optimization steps have been performed to improve the passivation of monolayer TMD defects/trap-states by chemical treatment\cite{KimACSN2017,BretscherACSN2021}, although the approach of Ref.\citenum{AmaniS2015} appears to be more straightforward. Furthermore, the impact of this chemical treatment on multilayer TMDs has yet to be tested.     

In this work, we report for the first time, a hysteresis-free ($\Delta n<10^{10}$\,cm$^{-2}$) electrical behaviour of a high quality heterostructure formed by graphene encapsulated in TFSI-treated multilayer tungsten disulfide (labeled here as Tr-WS$_2$). Specifically, we found that the hysteresis in the Tr-WS$_2$/Gr/Tr-WS$_2$ heterostructure can reach values of $\Delta n<5\cdot 10^{10}$\,cm$^{-2}$, which is the measured noise level, and at least a factor five lower than $\Delta n\sim 2.6\cdot 10^{11}$\,cm$^{-2}$ measured for a similar device but without the chemical treatment. We attribute the large hysteresis in this latter untreated heterostructure mostly to the charge traps present in the WS$_2$ used as a gate dielectric. We confirmed the validity of the TFSI treatment on the electrical behaviour of the Tr-WS$_2$/Gr/Tr-WS$_2$ heterostructure by also performing sweep rate and sweep range measurements, which display no change in $\Delta n$. Notably, the hysteresis is also unaffected by ageing after 120 days, suggesting the resiliency and persistence of the acid treatment in three devices. Moreover, the room temperature mobility of the Tr-WS$_2$/Gr/Tr-WS$_2$ heterostructure at carrier density of $\sim$10$^{12}$\,cm$^{-2}$ is as high as $\sim$6.2$\cdot$10$^{4}$\,cm$^{-2}$V$^{-1}$s$^{-1}$ with a $n$* of $\sim 10^{11}$\,cm$^{-2}$. Mobility is stable and almost flat even at higher carrier density $n\sim$3.8$\cdot$10$^{12}$\,cm$^{-2}$. The above results suggest that graphene encapsulated in chemically Tr-WS$_2$ could be an interesting candidate for integrated applications\cite{Sze2006a,Romagnoli2018a,Kang2008,Muench2019a}, especially in a context where the heterostructure is replicated with all-CVD grown materials.

\section{\label{Results}Results}

\begin{figure*}[htbp!]
\centerline{\includegraphics[width=180mm]{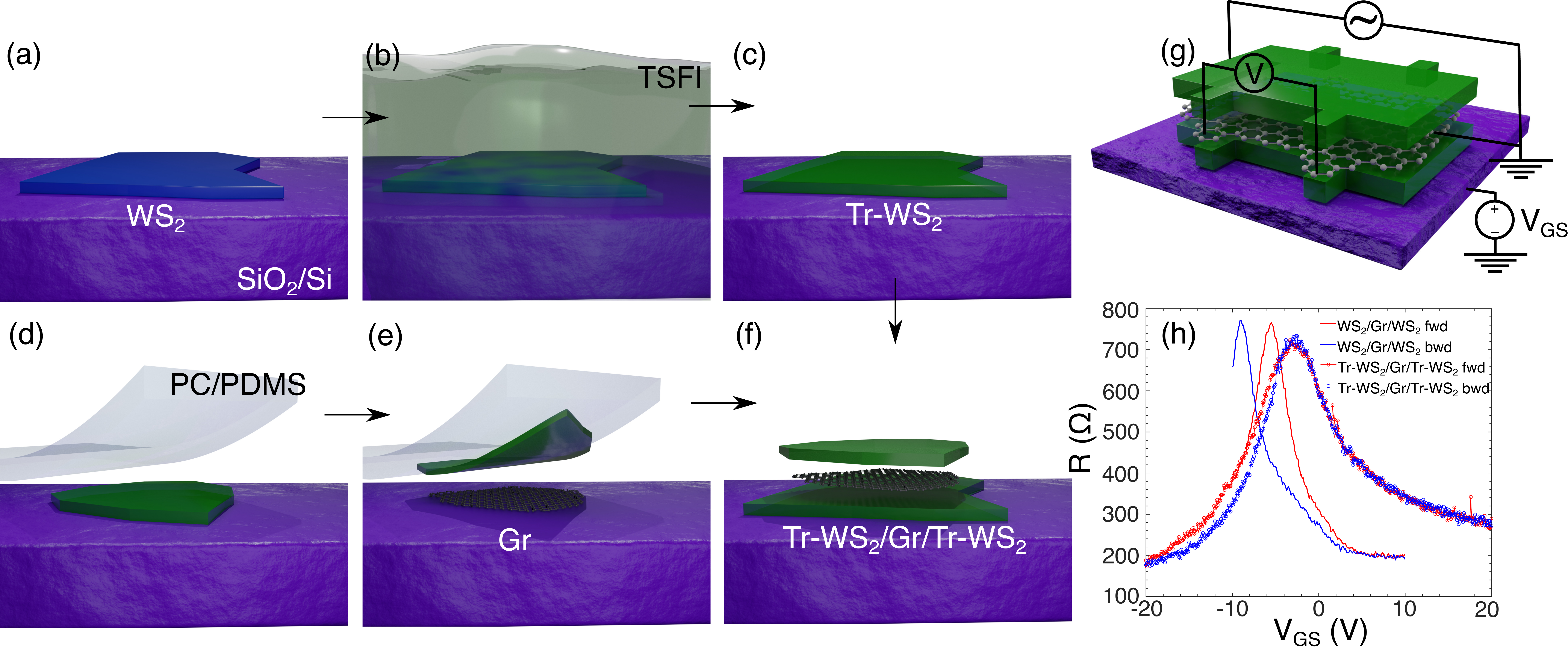}}
\caption{(a) Exfoliation of WS$_2$ on SiO$_2$/Si, (b) chemical treatment of the exfoliated flake using TFSI for $\sim$30 minutes and (c) dried WS$_2$ shown on SiO$_2$/Si, stored for later use. Processes (a-c) were performed inside a nitrogen glovebox. (d) A treated WS$_2$ flake on SiO$_2$/Si is picked up by a PC film and it is used as the top layer of the heterostructure, then (e) exfoliated graphene is picked up with the top WS$_2$. (f) The WS$_2$/Gr stack is then dropped on top of another treated WS$_2$ flake to form the final treated heterostructure (Tr-WS$_2$/Gr/Tr-WS$_2$). (g) Schematic front view of the measurement configuration of the Hall bar shaped Tr-heterostructure. (h) Resistance of graphene measured as a function of the back gate voltage V$_{\text{GS}}$ under a forward and backward sweep, comparing data between a treated and an untreated heterostructure.}
\label{fig:Fig1}
\end{figure*}

We start our work with fabricating an heterostructure made of Tr-WS$_2$/Gr/Tr-WS$_2$. The fabrication procedure starts with the micro-mechanical cleavage\cite{Novoselov2005b} of $\sim$6-25\,nm-thick WS$_2$ flakes and single layer graphene on 285\,nm SiO$_2$+Si chips. Optical microscopy\cite{Casiraghi2007} and Raman spectroscopy\cite{FerrPRL2006} were used to select the required flakes before proceeding with further processing, as depicted in Fig.\ref{fig:Fig1}. Next, we prepared the acid solution by dissolving TFSI in 1,2-dichlorethane (DCE) at a concentration of 5\,mg/ml. The exfoliated WS$_2$ flakes on SiO$_2$/Si substrates were immersed in the acid for about 30\,minutes and then blow dried using a nitrogen gun as shown in Fig.\ref{fig:Fig1}(a-c). All these processes were performed inside a nitrogen glovebox as suggested by Ref.\citenum{BretscherACSN2021}. In parallel, we prepared a polycarbonate (PC) stamp and placed it over a polydimethylsiloxane (PDMS) support on a glass slide, in order to be able to dry-transfer the exfoliated flakes, as reported by Ref.\citenum{PurdieNC2018}. We then started assembling the heterostructure, by first picking up a Tr-WS$_2$ flake from a SiO$_2$/Si substrate at 40°C, as shown in Fig.\ref{fig:Fig1}(d). Next, a single layer graphene is picked up using the stamp+Tr-WS$_2$, again at 40°C. Finally, the whole stack is dropped on the bottom Tr-WS$_2$ (Fig.\ref{fig:Fig1}(c)), followed by release of the PC film at 120°C \ref{fig:Fig1}(e-f) and dissolution of PC in chloroform lasting $\sim$10 minutes. The above heterostructure is then shaped in the form of an Hall bar (see Methods\ref{Methods}). In order to disentangle the role of the top and bottom WS$_2$ layers in the device hysteresis, we also prepared an Hall bar where the WS$_2$ layers were not treated with TFSI (WS$_2$/Gr/WS$_2$, untreated heterostucture). 

Fig.\ref{fig:Fig1}(g) is a schematic of our four probe measurement setup: a bias current is driven between the source and drain external contacts, whilst the voltage difference is read between the two inner probes. A back gate voltage is applied through the bottom (Tr-WS$_2$ or WS$_2$)+SiO$_2$ insulating layers. Fig.\ref{fig:Fig1}(h) shows the four probe resistance measurements performed on the treated and untreated heterostructures by sweeping the back gate voltage (V$_{\text{GS}}$). In order to evaluate the hysteresis, we performed forward (red) and reverse (blue) sweeps of the untreated (solid line) and treated (line with circles) heterostructures. The untreated heterostructure exhibits a considerable shift in the CNP, $\Delta$V$_{\text{CNP}}$, of $\sim$5V). This shift could be explained as follows: when we sweep the back gate voltage, depending on the polarity, electrons or holes transfer from the graphene to trap sites on the substrate and become trapped. On sweeping back the gate voltage, the trapped charges electrostatically dope graphene, which manifests in shifting its CNP\cite{WangACN2010}. In our heterostructure, hysteresis is caused by the charge trapping of carriers from graphene within the WS$_2$ vacancy sites\cite{Nan2014,LLi2019} and also at the interface with the SiO$_2$ surface\cite{WangACN2010}, with the former being orders of magnitude more dominant with respect to the latter\cite{QiuNC2013,TAndo1982,WangACN2010}. The heterostructure sample shows less than an order of magnitude hysteresis compared with the untreated heterostructure, with a $\Delta$V$_{\text{CNP}}$ of $\sim$0.2V. The charge trap density $\Delta n$ is then also a measure of the sample hysteresis and can be calculated from the $\Delta$V$_{\text{CNP}}$ as\cite{WangACN2010}:

\begin{equation}
\label{eq:Eq1}
\Delta n = \frac{C_{g}\cdot\Delta V_{\text{CNP}}}{2e}.
\end{equation}

\begin{figure*}[htbp!]
\centerline{\includegraphics[width=180mm]{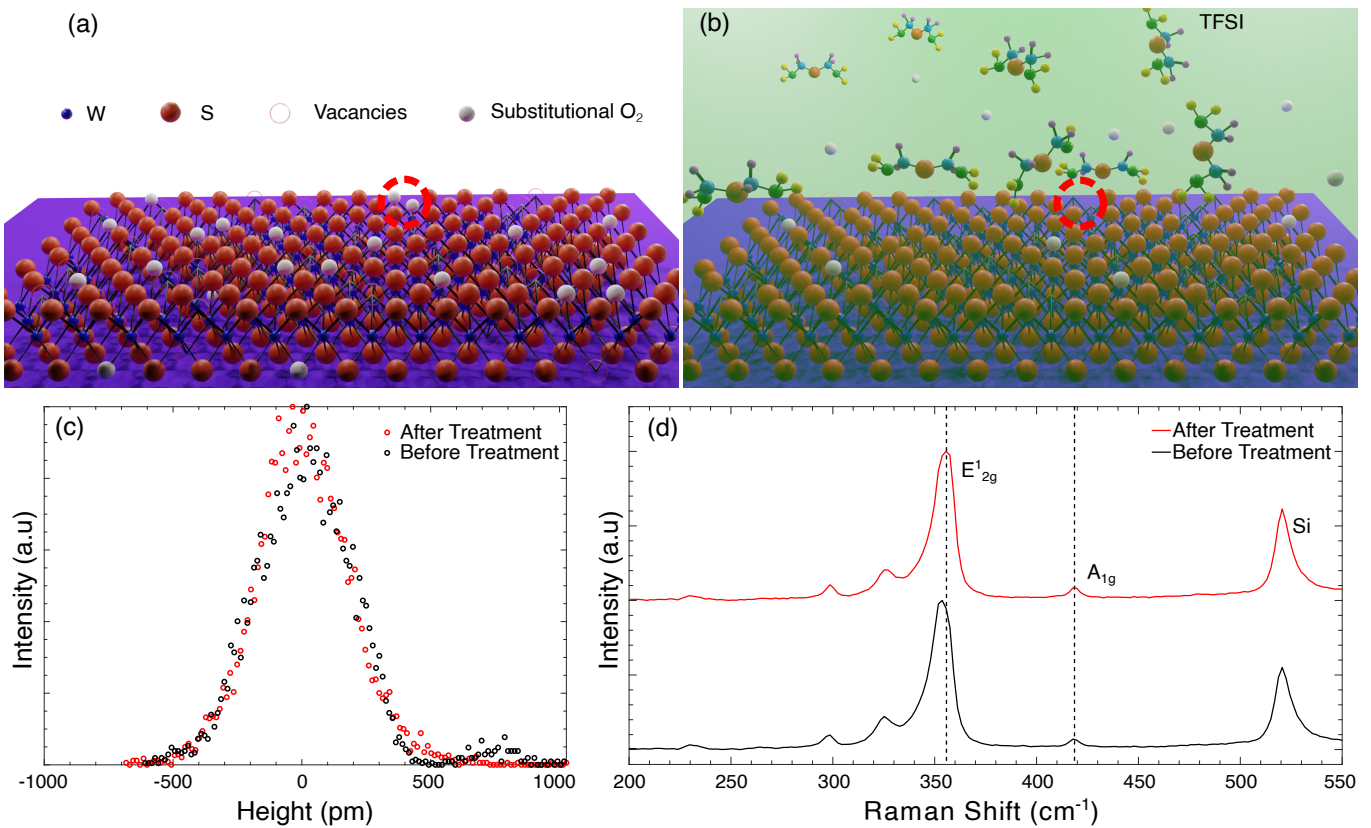}}
\caption{ (a) Graphical representation of the as-exfoliated WS$_2$ on SiO$_2$/Si with vacancies and substitutional oxygen occupancies (the red dashed line shows an area with two substitutional O$_2$ atoms) and (b) change of the surface after the treatment with TFSI (here the red dashed line represents how sulfur atoms have regained their place in the lattice instead of O$_2$ substitutional atoms). (c) AFM histogram comparison of the treated (red) and untreated (black) WS$_2$ on SiO$_2$/Si. (d) Raman spectrum of the treated (red) and untreated (black) WS$_2$ on SiO$_2$/Si.}
\label{fig:Fig2}
\end{figure*}

where C$_{g}$ is the gate capacitance calculated from:

\begin{equation}
\label{eq:Eq2}
C_{g} =\frac{\epsilon_0\cdot\epsilon_1\cdot\epsilon_2}{\epsilon_2\cdot d_1+\epsilon_1\cdot d_2}.
\end{equation}

As we gate graphene through two dielectrics, here $\epsilon_1$ is the relative permittivity of SiO$_2$ (3.9), $\epsilon_2$ is the relative permittivity of WS$_2$ (6.2, from Ref.\citenum{ALaturia2018}), $\epsilon_0$ is the vacuum permittivity, while $d_1$ and $d_2$ are the thickness values of SiO$_2$ and WS$_2$, respectively. We measured a trap density $\Delta n$ of $\sim 2.6\cdot 10^{11}$\,cm$^{-2}$ for the untreated heterostructure and $\sim 7\cdot 10^{9}$\,cm$^{-2}$ for the Tr-WS$_2$/Gr/Tr-WS$_2$ (treated heterostructure) by taking the absolute difference between the maximum values in resistance in forward and backward sweeps. However, we note that the resistance of the sample fluctuates, potentially masking the correct position of the CNP. A difference of 0.7\,V (corresponding to $\Delta n\sim 5\cdot 10^{10}$\,cm$^{-2}$) is indeed observed between the position of the CNP taken from the raw data as the maximum value of resistance and the same value taken while smoothing the data, instead. We then infer that a more conservative value of $\Delta n\sim 5\cdot 10^{10}$\,cm$^{-2}$ can be claimed as upper bound of hysteresis, which is certainly not overcome by the treated heterostructure. From now on we refer to this value as the noise level. This clearly indicates that the TFSI treatment played a crucial role in suppressing the defect sites, drastically decreasing the device hysteresis by almost one order of magnitude compared to the untreated WS$_2$/Gr/WS$_2$ heterostructure.

In order to be able to address the exhibited behaviour of the treated heterostructure in Fig.\ref{fig:Fig1}(h) one has to carefully analyse the existing problem of hysteresis in TMD-based devices and the origin of such effect. The predominant intrinsic defect types in as-exfoliated and as-grown TMDs are chalcogen vacancies, depicted as hollow spheres in Fig.\ref{fig:Fig2}\cite{Lin2016}. At ambient conditions, these vacancies are often filled with substitutional atoms, mainly oxygen\cite{Liu2016,Peto2018,Barja2019}, represented as white spheres in Fig.\ref{fig:Fig2}. Depending on their concentration, defects can considerably alter the electronic and optical properties of TMD-based devices\cite{Wang2018,ZHe2016}. Vacancies represent the main source of trapping/detrapping of charges\cite{Nan2014}, which in turn results in hysteresis effects\cite{HWang2018,HAn2020}, especially when a TMD is used as a dielectric. Many methods have been proposed to eliminate the substitutional atoms and/or ``passivate'' the vacancies with the original chalcogen atom species\cite{AmaniS2015,KimACSN2017,BretscherACSN2021,YLiu2022}. 

\begin{figure*}[htbp!]
\centerline{\includegraphics[width=180mm]{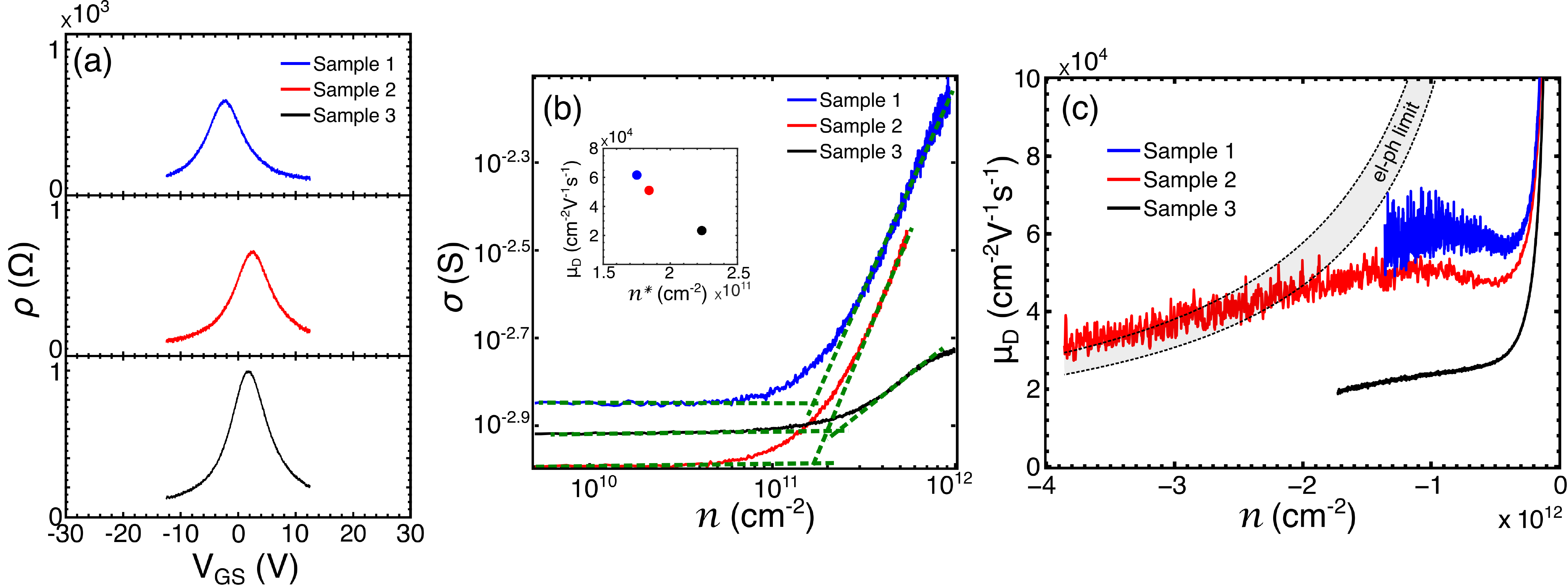}}
\caption{(a) Device resistivity ($\rho$) comparison of the TFSI-treated samples as a function of gate voltage V$_{\text{GS}}$. (b) Conductivity of the treated samples as a function of $n$ plotted in log scale; the inset plots $\mu_D$ as a function of $n$*. (c) Drude mobility ($\mu_D$) comparison of the TFSI-treated samples as a function of carrier density $n$. The grey shaded area with the black dotted borders indicates the electron-phonon limited mobility.}
\label{fig:Fig3}
\end{figure*}

In this work, we chose to perform a non invasive chemical approach as proposed in Ref.\citenum{AmaniS2015}. As schematized  by the red dashed circle in Fig.\ref{fig:Fig2}(a) and (b), the treatment with TFSI removes the substitutional oxygen atoms or any other absorbed contaminant and passivates the vacancy site\cite{AmaniS2015,BretscherACSN2021}. Performing a chemical treatment could theoretically deteriorate the morphological properties of the TMD. For this reason we scanned by atomic force microscopy (AFM) an area of 15\,$\mu$m$\times$15\,$\mu$m of the sample before and after the treatment (\ref{fig:Fig2}(c) and insets) and we plotted an histogram of the height profiles. Black and red circles represent height profiles of the samples acquired before and after the chemical treatment, respectively. On fitting these height distributions with a Gaussian, we obtain a standard deviation which points towards a roughness around $\sim$155\,pm and $\sim$145\,pm for the sample before and after the treatment, respectively. These AFM results suggest that the chemical treatment had produced no significant morphological change to the material surface. Raman characterisation was also performed using a Renishaw Invia Spectrometer with 532\,nm laser excitation and a 100$\times$ objective lens. Raman spectra of the WS$_2$ flake on SiO$_2$/Si before and after treatment are represented as the black and red lines in Fig.\ref{fig:Fig2}(d), respectively. On carefully examining the E$^{1}_{2g}$ mode at $\sim$350\,cm$^{-1}$ and the A$^{1}_{g}$ mode at $\sim$420\,cm$^{-1}$, we observe a slight blue-shift after the treatment, which has been associated with a reduction of the doping in the sample\cite{YLiu2022,Yang2017,Zhao2013}. This effect is in agreement with the removal of the substitutional oxygen and passivation of the vacancy sites by the TFSI treatment explained earlier. Results from AFM and Raman demonstrate the robustness of the TFSI treatment, and therefore open a path for the exploration of more complicated heterostructures.  

\begin{figure*}[htbp!]
\centerline{\includegraphics[width=180mm]{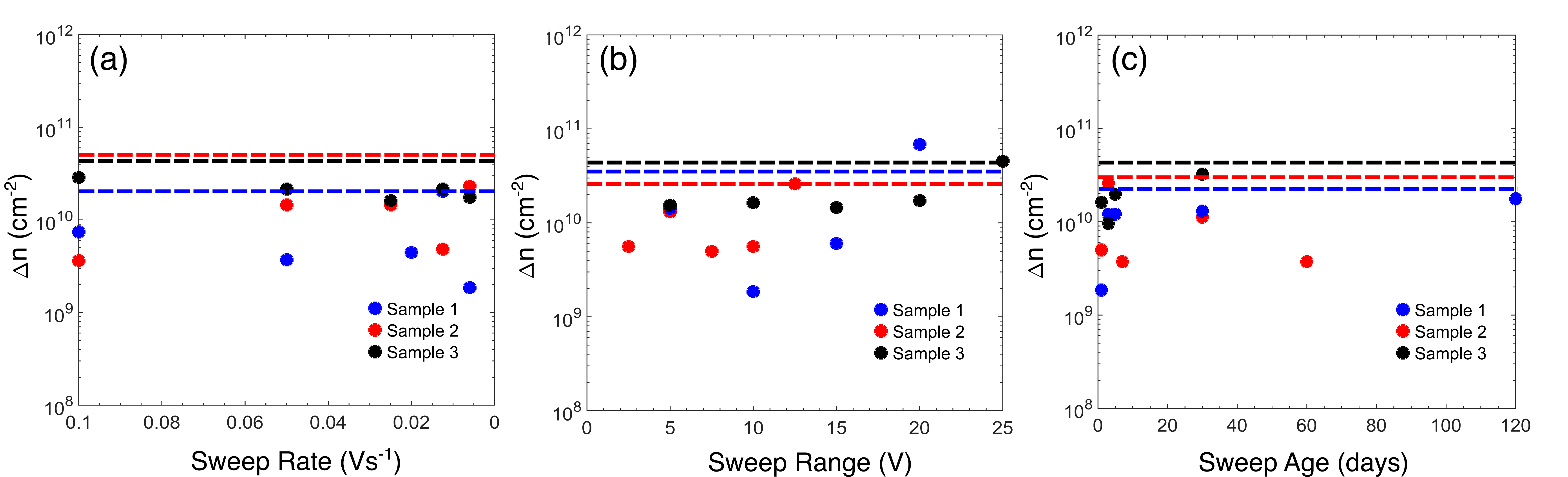}}
\caption{Systematic study of $\Delta n$ as a function of (a) sweep rate, (b) sweep range and (c) ageing over time for Sample 1 (blue), Sample 2 (red) and Sample 3 (black). The dotted lines in the plots are the noise levels of each sample.}
\label{fig:Fig4}
\end{figure*}

We then fabricated three different Hall bars out of Tr-WS$_2$/Gr/Tr-WS$_2$ heterostructures (which we name here as Sample 1, 2, 3, respectively), with thickness of the top and bottom WS$_2$ layers in the range $\sim$6-25\,nm. The length $L$ and width $W$ of the Hall bars are $L$=2\,$\mu$m-$W$=7\,$\mu$m for Sample 1, $L$=4\,$\mu$m-$W$=10\,$\mu$m for Sample 2, and $L$=1.5\,$\mu$m-$W$=2.5\,$\mu$m for Sample 3. We focused on the transport properties of such Hall bars measured at room temperature. Fig.\ref{fig:Fig3}(a) is the resistivity ($\rho$) of the treated Samples 1-3 measured while sweeping the backgate voltage V$_{\text{GS}}$. Here we use resistivity $\rho$ instead of resistance to have a more direct comparison among the three samples without accounting for geometrical differences among the Hall bars. The CNP in the three samples is positioned within ± 2.7\,V from zero gate bias, which correspond to variations in the Fermi level of graphene (at zero gate bias) in the three samples within ± 74\,meV\cite{ADas2008}, indicating a minimal doping. The $\rho$ values in Samples 1-3 are comparable, varying from $\sim$640-990\,$\Omega$ at CNP to $\sim$100-120\,$\Omega$ at high doping. Fig.\ref{fig:Fig3}(b) shows the measured conductivity of graphene at room temperature, plotted on a double log scale as a function of $n$. The carrier density $n$ is calculated from $n$ =  $C_{g}$(V$_{\text{GS}}-$V$_{\text{CNP}}$)/$e$\cite{PurdieNC2018}. $n^{*}$ can be extracted from this plot by noting the intersection point between the two quasi-linear sections of the conductivity plot as highlighted by the green dashed line. The $n^{*}$ for Sample 1, 2, and 3 are $\sim$1.7, $\sim$1.8 and $\sim$2.2$\cdot 10^{11}$\,cm$^{-2}$, respectively. The Drude mobility $\mu_{D}$ is calculated from $\mu_{D}$ = $\sigma$/($ne$), where $\sigma$ = 1/$\rho$ is the conductivity of graphene.  At technologically relevant charge carrier densities $n\sim$ 10$^{12}$\,cm$^{-2}$, Sample 1, 2 and 3, exhibit a $\mu_{D}$ of $\sim$6.2, $\sim$5.0 and $\sim$2.5$\cdot$10$^{4}$\,cm$^{-2}$V$^{-1}$s$^{-1}$, respectively, as shown in the inset of Fig.\ref{fig:Fig3}(b). The slight difference in the mobility among the samples, can be correlated to the $n^{*}$ and agrees well with the inversely linear relationship between $\mu_D$ and $n^{*}$\cite{Couto2014}. 

In general, at higher $n$ ($>10^{12}$\,cm$^{-2}$) carriers couple to acoustic phonons and the mobility degrades\cite{APrincipi2014,MAYamoah2017,Hwang2008b}. Fig.\ref{fig:Fig3}(c) is the plot of the $\mu_{D}$ as a function of the charge carrier density $n$. The gray shaded region with the black dotted borders illustrates the upper and lower limit of the electron-phonon limited mobility model from Ref.\citenum{Hwang2008b}, which are calculated assuming deformation potential (D) values of 18-20\,eV\cite{Hwang2008b}. We then decided to probe the sample with intermediate values of mobility up to a $n\sim$-3.8$\cdot10^{12}$\,cm$^{-2}$ and we recorded a factor $\sim$30\% drop in mobility compared to the maximum acceptable value measured at $n\sim$-1$\cdot10^{12}$\,cm$^{-2}$. This is a factor two lower compared to the mobility drop recorded in hBN-encapsulated graphene devices, which typically undergo a $\sim$70\% drop in mobility in the same doping range\cite{HuangNC2020,Wang2013g}. This behaviour is one crucial advantage of TMD-encapsulated graphene devices over hBN-encapsulated graphene devices, especially in the context of applications such as photodetectors or modulators operating at high doping. A comparison between the record-high mobility obtained with hBN-encapsulated graphene samples and the treated-TMD-encapsulated graphene samples used here would not be fair, as the reported values of the former refer to using hBN thick flakes of $\sim>$80\,nm \cite{PurdieNC2018, HuangNC2020}. In this work the thickness of both top and bottom WS$_2$ layers in all samples were kept below $\sim$25\,nm. We strived to contain the WS$_2$ thickness below $\sim$25\,nm so that the used materials would be as similar as possible, in terms of thickness, to those successfully grown in the research community by CVD\cite{JungSR2021,CunninghamJPC2016,SchramESSDERC2017}. While comparing our samples with those using hBN flakes of the same thickness, we obtain similar values of $\mu_{D}$\cite{PurdieNC2018}. 

In order to probe the reliability of the treatment and its durability in time, we monitored the $\Delta n$ in Samples 1-3 while changing the sweep rate, range and re-testing the samples over time. The voltage sweeping rate is very important as reducing the sweep rate gives charges sufficient time to become trapped in the defect site, resulting in higher hysteresis. Therefore, one should expect an increase in hysteresis, while reducing the sweep rate if traps are present in the Tr-WS$_2$. However, Fig.\ref{fig:Fig4}(a) shows that by reducing the sweep rates from 0.1\,Vs$^{-1}$ to 0.005\,Vs$^{-1}$ no significant change in the hysteresis values $\Delta n$ occurs: remarkably, all the samples show very low variations of $\Delta n$ below the noise level for any of the utilised sweep rates, confirming the high efficiency of the treatment in passivating the defect sites and hindering charge trap mechanisms. The $\Delta n$ of Samples 1-3 are lower than the noise level measured for each of the samples, indicated as color-matched dashed lines in Fig.\ref{fig:Fig4}(a). Next, we monitored the $\Delta n$ with respect to the V$_{\text{GS}}$ sweep range, as we should expect an increase in the device hysteresis with the increase in the voltage sweep range\cite{WangACN2010}. This is due to the fact that, by increasing the V$_{\text{GS}}$ sweep range, more and more carriers become available for transport, but they are also more likely to become trapped in the defect sites. In Fig.\ref{fig:Fig4}(b) we plot $\Delta n$ against various gate voltage ranges, where, for example, a V$_{\text{GS}}$ sweep from -10\,V to 10\,V corresponds to a voltage range of 10\,V, considering the CNP being close to 0\,V. The measurement range of each sample is limited by the WS$_2$ and SiO$_2$ breakdown voltages, assumed to be $\sim$0.1\,MVcm$^{-1}$\cite{YFan2017} and $\sim$15\,MVcm$^{-1}$\cite{McPherson2003}, respectively. For these measurements the sweep rate is kept constant at 0.0125\,Vs$^{-1}$. Samples show almost no dependence on the sweep range with $\Delta n$ kept consistently below $\sim 5\cdot 10^{10}$\,cm$^{-2}$. However, when the V$_{\text{GS}}$ sweep range was increased to 20\,V for Sample 1, having 6\,nm bottom Tr-WS$_2$ thickness, we start to notice an increase in $\Delta n$ to $\sim 7\cdot 10^{10}$\,cm$^{-2}$. This particular behaviour could be attributed to the onset of the dielectric breakdown of the the bottom Tr-WS$_2$ flake\cite{YFan2017}. 

Next, the $\Delta n$ of the treated heterostructures is monitored at regular time intervals to check the integrity of the TFSI treatment over time. For these measurements we also adopted a sweep rate of 0.0125\,Vs$^{-1}$. Samples were measured until 120 days from the fabrication date and data are shown in Fig.\ref{fig:Fig4}(c). All the three samples show again no dependency towards the ageing, confirming the resilience of the chemical treatment. It is worth mentioning that the samples were stored in ambient lab atmosphere for the entire duration of this experiment. In any of the measurement conditions $\Delta n$ has never overcome $\sim 5\cdot 10^{10}$\,cm$^{-2}$. To highlight the significance of the hysteresis treatment when measuring a Tr-WS$_2$/Gr/Tr-WS$_2$ heterostructure, we outline the noise level of the Samples 1-3 in each of the graphs of Fig.\ref{fig:Fig4}, and all the data-points are below such values of noise, proving that the hysteresis is indeed minimal.    

\section{\label{Conclusion}Conclusion}
In conclusion, we have demonstrated that TFSI-treated WS$_2$ can be an excellent choice for encapsulating graphene and also performs well as gate dielectric. The TFSI treatment is highly stable, resilient and it drastically reduces the hysteresis in high mobility WS$_2$/Gr/WS$_2$ heterostructures. This could be due to the removal and replacement of substitutional atoms responsible for charge traps, as suggested by a Raman analysis and transport measurements. The hysteresis remains well under $\Delta n\sim 5\cdot 10^{10}$\,cm$^{-2}$, even when the sample is subjected to large sweep rates, range and ageing tests. This hysteresis value is at least five times smaller than that measured in an untreated heterostructure. In one treated heterostructure we achieved a mobility as high as $\sim$6.0$\cdot$10$^{4}$\,cm$^{-2}$V$^{-1}$s$^{-1}$. Due to the weak influence of surface acoustic phonons, the mobility of the treated heterostructures are almost flat up to $n\sim 3.8\cdot10^{12}$\,cm$^{-2}$. These results suggest that TMD/Gr/TMD heterostructures could be adopted in advanced optoelectronic applications requiring low hysteresis and high mobility at high carrier concentrations.

\section{\label{Methods}Methods}

\textit{Materials}: WS$_2$ highly crystalline van der Waals (vdW) crystals were synthesized through the self-flux technique. In the first step, the stoichiometric ratio of elemental puratronic grade (99.9999\% pure) purity tungsten powder and sublimation purified sulfur powder (99.9999\% pure) were mixed and sealed under 10$^{-6}$\,Torr inside a quartz ampoule. The powder was heated at 950°C for 5 days to create vdW powders. In the second step, these pre-reacted powders were resealed in a quartz ampoule with extra sulfur to reach 1:2 stoichiometry. In the second step, all the ingredients were kept at a high-temperature zone (1040°C) with a temperature drop of 50°C to initiate the vapor transport and realize vdW WS$_2$ crystals after 7 weeks. The resultant crystals measured 2-6\,mm in size. Overall, the crystal appearance and spectroscopic signatures were superior to chemical vapor transported crystals. Single layer graphene was exfoliated from highly-oriented pyrolytic graphite (HOPG) commercially bought from HQ Graphene labelled as natural graphite. TFSI (product number 15220) and 1,2-dichlorethane (DCE) (product number 284505) were bought from Sigma-Aldrich. 

\textit{Fabrication}: Lithography was performed by using an Inspect F50 RAITH Elphy Plus system and a poly(methyl methacrylate) (PMMA) mask. The heterostructure was then dry-etched (reactive ion etching) using a trifluoromethane/oxygen (CHF$_3$/O$_2$) mixture of gases with a flow rate of 40/4\,sccm in a Plasmalab System 100 from Oxford Instruments. Cr was deposited thermally with thickness of 5\,nm using a Lesker LAB18 evaporator, while Au was deposited thermally with the same instrument and with a 50\,nm thickness. The evaporation was followed by an overnight lift-off in acetone, dipping in isopropanol and nitrogen blow-drying.

\textit{Electrical Transport Measurement}: transport measurements were performed using a Lakeshore cryogenic probe station at a pressure of $\sim10^{-3}$\,mbar with a lock-in amplifier from Stanford Research Systems SR810. A bias of 100\,nA was applied between the source and drain electrodes with a lock-in modulation frequency of 373\,Hz. The gate voltages were applied using a Keithley 2460 sourcemeter. 

\section{\label{Ackn}Acknowledgements}
The authors acknowledge Matteo Ceccanti for graphical support and Francisco Bernal Texca for his help towards setting up the fabrication steps inside the glovebox. F.H.L.K. acknowledges financial support from the Government of Catalonia trough the SGR grant and from the Spanish Ministry of Economy and Competitiveness through the “Severo Ochoa” Programme for Centres of Excellence in R\&D (CEX2019-000910-S), support by Fundacio Cellex Barcelona, Generalitat de Catalunya, through the CERCA program, and the Mineco Grants Plan Nacional (FIS2016-81044-P) and the Agency for Management of University and Research Grants (AGAUR) 2017 SGR 1656.
Furthermore, the research leading to these results has received funding from the European Commission in the Horizon 2020 Framework Programme under Grant Agreements Nos. 785219 (Core2) and 881603 (Core3) of the Graphene Flagship. This work was also financially supported by the German Science Foundation (DFG) within the priority program FFlexCom Project “GLECS” (Contract No. NE1633/3). D.D.F. acknowledges funding from the “Severo Ochoa” Excellence Programme. S.T acknowledges support from DOE-SC0020653 (materials synthesis), NSF DMR-2206987 (magnetic characterization), NSF ECCS 2052527 (electrical optimization), and NSF DMR 2111812 (optical optimization).

\end{document}